\documentstyle[amsfonts,aps,preprint]{revtex}

\begin{document}
\draft
\title{On $(b,c)$-system at finite temperature in thermo field approach}
\author{M. C. B. Abdalla, A. L. Gadelha and I. V. Vancea \thanks{%
On leave from Babes-Bolyai University of Cluj.}}
\address{{\em Instituto de F\'{\i}sica Te\'{o}rica, Universidade Estadual 
Paulista}\\
{\em Rua Pamplona 145, 01405-900, S\~{a}o Paulo, SP, Brazil}}
\maketitle

\begin{abstract}
We construct the finite temperature field theory of the two-dimensional
ghost-antighost system within the framework of thermo field theory.
\end{abstract}

\date{\today }

Over the years, thermo field theory has provided to be an useful tool for
investigating quantum systems at finite temperature
\cite{tu,mnumm,io,umt,hu}. Beside its various applications to quantum field
theory \cite{umt,ad}, the thermal field approach was used in solving the
master equation in quantum optics \cite{qo}, studying the quantum stochastic
differential equation \cite
{qs}, quantum transport theory \cite{qtt} and formulating the
non-equilibrium density matrix as well as the entropy operator in the
operatorial approach \cite{ms}. More recently, it has been shown that
quantum deformations of the envelopping algebras are related to coherent
state counterpart in thermo field theory \cite{crv} and that the thermo
field algebra is incorporated in the $q$-deformation of Weyl-Heisenberg
algebra $h_{q}\left( 1\right) $ in the bosonic case \cite{msv}.
In this note we apply the thermo field construction to a two-dimensional
field theory composed by a ghost and an anti-ghost, also know as a
$(b,c)$-system. The main motivation for studying the $(b,c)$-system at finite
temperature is that in conformal two-dimensional quantum field theories the
use of the ghost and the anti-ghost fields is necessary in order to fix the
conformal invariance. At finite temperature, the conformal symmetry might
still survive, at least at low temperatures. Thus, in order to correctly
compute the physical states at non-zero temperature, one has to introduce
the $(b,c)$-fields. This may be the case of critical field theories \cite
{mcba} or string theories which have already been studied at finite
temperature but within the path integral formalism (for review see, for
example, \cite{lag}). An vertex algebra point of view on the $(b,c)$-system
was recently presented in \cite{efh}.

Let us consider the ghost field $c(z)$ and the anti-ghost field $b\left(
z\right) $ associated to a two-dimensional conformal symmetry in
two-dimensions. The action of this system is given by the following relation
\cite{fms} 
\begin{equation}
S\sim \int d^{2}z\left[ b\overline{\partial }c+\overline{b}\partial
\text{\/} \overline{c}\right]. \label{action}
\end{equation}
In what follows we consider only the fields with odd Grasmann parity, but
the results can be straightforwardly applied to even parity fields.
Since $ b\left( z\right) $ and $c\left( z\right) $ are holomorphic, the
following expansions hold 
\begin{equation}
b\left( z\right) =\sum\limits_{n}b_{n}z^{-n-2},\qquad c\left( z\right)
=\sum_{n}c_{n}z^{-n+1},  \label{exp}
\end{equation}
and the coefficients $b_{n}$ and $c_{n}$ satisfy the following algebra 
\begin{equation}
\left\{ c_{n},b_{m}\right\} =\delta _{n+m,0},\qquad \left\{
c_{n},c_{m}\right\} =\left\{ b_{n},b_{m}\right\} =0.  \label{alg}
\end{equation}
The energy-momentum tensor $T\left( z\right) $ and the ghost curren
number $ j\left( z\right) $ can be expanded in powers of $z$ with the
coefficients $L_{n}$ and $j_{n}$, respectively. In quantum theory the
operators $L_{n}$ satisfy a central extension of the Virasoro algebra
with the central charge $c_{g}=-26$. In the bosonic string theory $c_{g}$
compensates for the central charge of the bosonic degrees of freedom
\cite{fms}. The vacuum of the theory is degenerate with respect to the
eigenvalues of the operator $j_{0}$. The $\left| q\right\rangle $-vacuum 
is defined by the following relations
\begin{eqnarray}
b_{n}\left| q\right\rangle &=&0,\quad n>q-2,  \label{vac} \\
c_{n}\left| q\right\rangle &=&0,\quad n\geq -q+2,  \nonumber
\end{eqnarray}
and it is an eigenvector of both $j_{0}$ and $L_{0}$ \cite{fms}.

In order to construct the finite temperature counterpart of the
$(b,c)$-system following the thermo field prescription \cite{tu}, one has
to double the system. The identical copy of it will be denoted in what
follows by a tilde. The operators $\widetilde{b}_{n}$ and
$\widetilde{c}_{n}$ are taken to satisfy (\ref{alg}). Moreover, the two
copies are considered to be completely independent. The Hilbert space of
the extended system is a direct product of the two Hilbert spaces
${\cal H}\otimes \widetilde{{\cal H}}$. The next step is to construct a
Bogoliubov operator $G_{n}\left( T\right) $ for each oscillation mode that
acts on the total Hilbert space and maps each oscillator into its image at
temperature $T$.

The main difficulty in implementing this program is two-fold: on one hand,
due to the definition (\ref{vac}) of the vacuum, the same operator can be
either a creation or an annihilation operator for different $q$'s. On the
other hand, from the algebra (\ref{alg}) results that for each $q$ there are
two types of oscillators: the first one has $c_{-n}$ as creation operator an 
$b_{n}$ as annihilation operator for $n\geq q-1$. The second one has $b_{-n}$
as creation operator and $c_{n}$ as annihilation operator for $n\geq -q+2$.
The two operators differ, in principle, from each other because the fields
$b$ and $c$ have opposite ghost numbers. The Bogoliubov operators for the
two types of oscillators are given by \cite{tu} 
\begin{eqnarray}
G_{n}^{1} &=&-i\theta _{n}\left( T\right) \left( \widetilde{b}%
_{n}b_{n}-c_{-n}\widetilde{c}_{-n}\right) ,  \nonumber \\
G_{n}^{2} &=&-i\theta _{n}\left( T\right) \left( \widetilde{c}%
_{n}c_{n}-b_{-n}\widetilde{b}_{-n}\right) ,  \label{gop}
\end{eqnarray}
where the superscripts 1 and 2 denote the type of oscillator and $\theta
_{n}\left( T\right) $ is a real function of $T$. This function also depends
on the statistics of the oscillator. In the present case, the two types of
oscillators have the same statistics. Using the algebra (\ref{alg}), it is
easy to see that $G_{n}^{1}$ and $G_{n}^{2}$ commute for all the values of
$ n $ for the same type of oscillators and, also, commute with each other.
Moreover, for each $n$ they obey the following ${\Bbb Z}_{2}$ symmetry
$G_{-n}^{1}=G_{n}^{2}$ if $\theta _{n}\left( T\right) $ remains unchanged.
This is true if the frequencies of $n$ and $-n$ oscillators satisfy $\omega
_{n}=\omega _{-n}$, what we always assume. Due to this particular structure
of the $(b,c)$-system, there are two possibilities to map each oscillator
into its counterpart at finite temperature. Both of these mappings are
allowed by the thermo field construction. Nevertheless, just one of them is
fully compatible with the structure of the $\left( b,c\right) $-system. We
shall analyze both cases in what follows.

Let us denote by $\left| \left. q\right\rangle \hspace{-0.03in}\right\rangle 
$ the $q$-vacuum of the extended system at zero temperature and by $A_{n}$
any of the operators $b_{n}$, $c_{n}$, $b_{-n}$ and $c_{-n}$. The first case
to be discussed is when the $q$-vacuum transforms under the action of all
$G$
\begin{equation}
\left| \left. q\left( T\right) \right\rangle \hspace{-0.03in}\right\rangle
=\prod_{k=-\infty }^{\infty }e^{iG_{k}^{1}\left( T\right) }\prod_{s=-\infty
}^{\infty }e^{iG_{s}^{2}\left( T\right) }\left| \left. q\right\rangle 
\hspace{-0.03in}\right\rangle ,  \label{vact}
\end{equation}
where $\left| \left. q\left( T\right) \right\rangle
\hspace{-0.03in}\hspace{-0.03in}\right\rangle $ denotes the $q$-vacuum at
the temperature $T$. The operator $A_{n}\left( T\right) $ is given by
\begin{equation}
A_{n}\left( T\right) =e^{iG_{n}^{2}\left( T\right) }e^{iG_{n}^{1}\left(
T\right) }A_{n}e^{-iG_{n}^{1}\left( T\right) }e^{-iG_{n}^{2}\left( T\right)
}.  \label{opt}
\end{equation}
Using the commutation properties of $G_{n}^{1}\left( T\right) $ and $%
G_{n}^{2}\left( T\right) $, it is easy to show that $\left| \left. q\left(
T\right) \right\rangle \hspace{-0.03in}\right\rangle $ satisfy the following
relations 
\begin{eqnarray}
b_{n}\left| \left. q\left( T\right) \right\rangle \hspace{-0.03in}%
\right\rangle &=&0,\quad n\geq q-1,  \label{qvact} \\
c_{n}\left| \left. q\left( T\right) \right\rangle \hspace{-0.03in}%
\right\rangle &=&0,\quad n\geq -q+2,  \nonumber
\end{eqnarray}
and similar ones for tilde operators which transform as in (\ref{opt}). The
above relations show that $\left| \left. q\left( T\right) \right\rangle 
\hspace{-0.03in}\right\rangle $ is a $q$-vacuum of the system at finite
temperature. Indeed, one can show that 
\begin{eqnarray}
\left\{ c_{n}\left( T\right) ,b_{m}\left( T\right) \right\} &=&\left\{
c_{n},b_{m}\right\} =\delta _{n+m,0},  \nonumber \\
\left\{ c_{n}\left( T\right) ,c_{m}\left( T\right) \right\} &=&\left\{
b_{n}\left( T\right) ,b_{m}\left( T\right) \right\} =0.  \label{algt}
\end{eqnarray}
Therefore, the oscillator structure is maintained at finite temperature.
Since the fields $b$ and $c$ are scalar fields, for $T\neq 0$ we have the
following relations defining $b\left( T\right) $ and $c\left( T\right) $%
\begin{equation}
b\left( T\right) =\sum\limits_{n}b_{n}\left( T\right) z^{-n-2},\qquad
c\left( T\right) =\sum_{n}c_{n}\left( T\right) z^{-n+1},  \label{ft}
\end{equation}
from which we conclude that the action has the same form as (\ref{action})
with the appropriate fields inside. Consequently, the energy-momentum tensor
and the ghost number current have formally the same expressions. From these
considerations we deduce that the conformal symmetry is not broken by
transformations (\ref{vact}) and (\ref{opt}). We can explicitly check this
conclusion by expanding the energy-momentum tensor and the ghost number
current in powers of $z$ and then expressing $L_{n}\left( T\right) $ and $%
j_{n}\left( T\right) $ in terms of $b\left( T\right) $ and $c\left( T\right) 
$. Then, since the latter set of operators satisfies the algebra (\ref{algt}%
), the algebra of $L_{n}\left( T\right) $ and $j_{n}\left( T\right) $ is
identical to that of $L_{n}$ and $j_{n}$.

Let consider now the second possibility. In this case, we focus our
attention on a specific $q$-vacuum $\left| \left. q\right\rangle \hspace{%
-0.03in}\right\rangle $. By acting with $G$-operator of either type one or
type two, we can transform $\left| \left. q\right\rangle \hspace{-0.03in}%
\right\rangle $ into the following vectors at finite temperature\footnote{%
In expressions (\ref{twovact}) we intoduce the modulus operation to avoid
ambiguities. Note that in the first case studied these ambiguities do not
take place for the reason that we have a combination of $G$-operators beside
to the fact that the ${\Bbb Z}_{2}$ symmetry is present. All of these operators
commute among themselves.} 
\begin{eqnarray}
\left| \left. q\left( T\right) \right\rangle \hspace{-0.03in}\right\rangle
_{1} &=&\prod_{k=q-1}^{\infty }e^{iG_{\left| k\right| }^{1}\left( T\right)
}\left| \left. q\right\rangle \hspace{-0.03in}\right\rangle ,  \nonumber \\
\left| \left. q\left( T\right) \right\rangle \hspace{-0.03in}\right\rangle
_{2} &=&\prod_{s=-q+2}^{\infty }e^{iG_{\left| s\right| }^{2}\left( T\right)
}\left| \left. q\right\rangle \hspace{-0.03in}\right\rangle .
\label{twovact}
\end{eqnarray}
If we denote by $A_{n}$ the operators $c_{-n}$ and $b_{n}$ for $n\geq q-1$
and by $B_{n}$ the set $b_{-n}$ and $c_{n}$ for $n\geq -q+2$ we have the
following transformations for the two types of oscillators
\begin{eqnarray}
A_{n}\left( T\right) &=&e^{iG_{\left| n\right| }^{1}\left( T\right)
}A_{n}e^{-iG_{\left| n\right| }^{1}\left( T\right) },  \nonumber \\
B_{n}\left( T\right) &=&e^{iG_{\left| n\right| }^{2}\left( T\right)
}B_{n}e^{-iG_{\left| n\right| }^{2}\left( T\right) },  \label{twoopt}
\end{eqnarray}
and similar relations for tilde operators. It is easy to verify that $\left|
\left. q\left( T\right) \right\rangle \hspace{-0.03in}\right\rangle _{1}$
and $\left| \left. q\left( T\right) \right\rangle \hspace{-0.03in}%
\right\rangle _{2}$ are $q$-vacua for type one and type two oscillators
respectively. However, once that various $q$'s are considered, the same
operator should transform under $G^{1}$ and $G^{2}$ simultaneously. That
spoils the construction. Therefore, (\ref{twovact}) and (\ref{twoopt}) are
not compatible with the multiple-vacuum structure of $(b,c)$-theory.
Nevertheless, it might be worthwhile to consider transformations of the type
(\ref{twovact}) and (\ref{twoopt}) under which a $q$-vacuum degenerates.
This might be useful in study of theories that emphasize a certain vacuum,
as could be the case of string theory in which the states are constructed
from the $SL\left( 2,{\Bbb R}\right) $-invariant vacuum.

The drawback of the transformations (\ref{twovact}) and (\ref{twoopt}) is
due to the presence of two types of oscillators and, apparently, it is
removable if one combines the two oscillators into a single one. That this
is not the case we shall prove by a counter example. Let us consider the
$0$-vacuum $\left| \left. q=0\right\rangle \hspace{-0.03in}\right\rangle$ and
let us denote by $\left( C_{n}^{\dagger },C_{n}\right) $ and $\left(
D_{n}^{\dagger },D_{n}\right) $ the first and second types of oscillators,
respectively. It is possible to combine $C$-operators and $D$-operators by
defining linear combinations
\begin{eqnarray}
E_{n} &=&c\left( n,0\right) C_{n}+d\left( n,0\right) D_{n},  \nonumber \\
E_{n}^{\dagger } &=&c^{*}\left( n,0\right) C_{n}^{\dagger }+d^{*}\left(
n,0\right) D_{n}^{\dagger },  \label{lincomb}
\end{eqnarray}
for $n\geq 2$, where $c\left( n,0\right) $ and $d\left( n,0\right) $ are
complex numbers depending on $n$ and $q=0$ and $*$ denotes the complex
conjugation. For $n=0,1$ the above combination should reduces to $C_{n}$ and 
$C_{n}^{\dagger }$ since $D_{n}$ and $D_{n}^{\dagger }$ are not defined for
these values of $n$. Thus, for $n=1$ the coefficients in (\ref{lincomb})
should satisfy
\begin{equation}
c^{*}\left( 1,0\right) =d\left( 1,0\right) =0,\quad d^{*}\left( 1,0\right)
=c\left( 1,0\right) =1,  \label{contr}
\end{equation}
which represents a contradiction. For $q\geq 2$ or $q\leq -2$, the ``gap''
in which a combination of type (\ref{lincomb}) reduces to one operator only
increases and also does the number of coefficients that should satisfy
relations like (\ref{contr}). Other linear combinations of $C_{n}$ and
$ D_{n} $ operators are possible, but all of them lead to contradictions like
the above ones.

To conclude, we have constructed in this note the finite temperature
counterpart of the $(b,c)$-system using the thermo field approach. We have
shown that there are two possibilities to obtain the finite temperature
vacuum and creation and annihilation operators. In the first case, all
$q$-vacua have a $T\neq 0$ correspondent. The transition does not break the
conformal invariance of the ghost-antighost system. In this case, the
$(b,c)$-system can be used to fix the gauge of a conformal invariant field
theory at finite temperature. The thermodynamics should be constructed from
the physical space of the matter and ghost fields. In the second case, the
transformation of $q$-vacua at $T=0$ can be defined only for one vacuum at a
time. Thus, the Bogoliubov operators act as transformation on isolate vacuum
and map each vacuum into two states at $T\neq 0$. Each of these states
represent a $q$-vacuum for the corresponding type of oscillator. The present
analysis might, hopefully, be useful in studying the conformal invariant
field theories at finite temperature in the framework of thermo field
theory. For example, it would be interesting to compare the results obtained
through this method with the ones given by the path integral approach in the
case of string theories.

We want to thank Hebe Queiroz Pl\'{a}cido for useful discussions. A.L.G.
acknowledgs CAPES for his fellowship. The work of I.V.V. was supported by
a FAPESP postdoc fellowship.

\end{document}